Evan Solomonides and Yervant Terzian

Cornell University

## A Probabilistic Analysis of the Fermi Paradox

### Abstract

The Fermi paradox uses an appeal to the mediocrity principle to make it seem counterintuitive that humanity has not been contacted by extraterrestrial intelligence. A numerical, statistical analysis was conducted to determine whether this apparent loneliness is, in fact, unexpected. An inequality was derived to relate the frequency of life arising and developing technology on a suitable planet in the galaxy; the average length of time since the first broadcast of such a civilization; and a constant term. An analysis of the sphere reached thus far by human communication was also conducted, considering our local neighborhood and planets of particular interest. We clearly show that human communication has reached a sphere with a radius of roughly 80 light years, and has not reached a number of stars and planets adequate to expect an answer. These analyses both conclude that the Fermi paradox is not, in fact, unexpected. By the mediocrity principle and numerical modeling, it is actually unlikely that the Earth would have been reached by extraterrestrial communication at this point. We predict that under 1% of the galaxy has been reached at all thus far, and we do not anticipate to be reached until approximately 50% of stars/planets have been reached. We offer a prediction that we should not expect this until at least 1,500 years in the future. Thus the Fermi paradox is not a shocking observation- or lack thereof- and humanity may very well be contacted within our species' lifespan (we can begin to expect to be contacted 1,500 years in the future).

### Introduction

"Our galaxy should be teeming with civilizations, but where are they?" (SETI, 1999). This unsettling question is known as the Fermi Paradox, and it has long puzzled astronomers. With over 200 billion stars in the Milky Way galaxy, and $1/2.4$ that many earthlike planets (Terzian and Hohlfeld, 1976), there are more places that life could have arisen than we could ever look. The Drake equation, which uses several parameters to offer a prediction as to the number of intelligent, communicating civilizations in the galaxy, almost invariably returns a number greater than 1, no matter how skeptically the parameters are chosen. The mediocrity principle (the idea that we are not even remotely special in the perspective of the universe/galaxy around us) tells us that whatever natural processes gave rise to life on our planet should be extraordinarily common, and should have happened in enumerable places all throughout the galaxy. Yet we are, to the best of our knowledge thus far, utterly alone. How can this be?

In this paper, an attempt is made to explain the Fermi paradox using a truth which is widely known but poorly understood: The galaxy is vast. And even though communications would travel at the speed of light, any civilizations attempting communication are sending their greetings across an almost unfathomably large distance. Even at the speed of light, these distances are daunting, and would take decades, centuries, and millennia to cross.

To provide a perspective on the vastness of the galaxy even compared to light-speed communication, an analysis is conducted of how far human signals have reached into the galaxy. The sphere of locations in space that human communication has reached (a sphere roughly 80 lightyears in radius) is analyzed in terms of the number of stars, planets, and earthlike planets it contains, and the portion of the Milky Way that it occupies. The negligible fraction of the galaxy that has been reached is offered as an intuitive explanation for the Fermi paradox.

A probabilistic analysis is conducted to determine how surprising it is that we have not been contacted yet. An equation is derived relating the length of broadcasting histories in the galaxy, the frequency of life occurring, and a constant. This equation is used to further remove the mystical nature of the Fermi paradox.

The analyses conducted make the Fermi paradox more of a statistically predictable result than a counterintuitive mystery.

## Analysis of Human Communication

We will take as a case study (and our only example upon which to offer predictions) an analysis of human broadcasts. The content of our broadcasts will be discussed, as well as the sphere which our communications have reached. Information determined here will be used to offer statistical predictions, because we have no other data point to make such predictions.

## History of Human Communication

The very first human signal powerful enough to leave the atmosphere and spread throughout our local stellar neighborhood was Adolf Hitler's voice (Trawick, 2011). The broadcast of his commentary of the 1936 Berlin Olympic Games has been moving away from Earth at the speed of light for roughly 80 years, and serves as the leading edge of a sphere of nearly-constant signals since. Though a handful of these signals have been designed to be picked up by extraterrestrial intelligence (SETI signals are mainly mathematical sequences that any civilization advanced enough to have built a radio receiver would recognize, i.e. Fibonacci, the prime numbers, the squares, all broadcast in binary), the vast majority would be indecipherable. This is because an alien civilization would need to first decode binary into sound (and figure out our tone encryption method) or video (with very specific, inconsistent formats), and if they could somehow do that, they would then need to decode the resulting 3,000 human languages (Anderson, 2012) into something they could parse successfully.

Though the messages we broadcast and intend for alien reception are designed to be easily interpreted, another civilization would have no way of distinguishing these from the overwhelming majority of human signals that would be identifiable as artificially-produced, but otherwise completely meaningless to any extraterrestrials, no matter how advanced. They simply would not have our particular data-compression algorithms, file formats, or languages. So the leading edge of our communication is Adolf Hitler speaking about the "inherent inferiority of non-Aryan athletes," and this is followed by decades and decades of what would be to other civilizations nothing but static. These broadcasts would signal that we are here and broadcasting, but would do very little else until the first SETI signals arrive 24 years later (SETI, 2016), if they're still bothering to try to decipher what it is we're sending out. Considering how we started broadcasting, perhaps it is best that the rest of the galaxy would have such a hard time figuring out what we're saying.

## The Sphere of Human Broadcasts

Almost all human broadcasts would thus be random, meaningless static to another civilized, technologically-advanced species. However, the universe does not naturally produce random, meaningless static. Any naturally-occurring signal is periodic and predictable, or otherwise identifiable, and has a rationally-explained, identifiable natural source (Tretkoff and Chodos, 2000). Therefore, our signals, however meaningless they may be to another civilization, would at least be a beacon indicating that we are here and ready to talk. We are left to hope that any sufficiently advanced civilization would be curious enough to try to continue the conversation.

It is known that any signal strong enough to leave Earth's atmosphere will propagate outwards in a sphere at the speed of light. Thus our communications have reached every star within about 80 lightyears of our sun. It is known that there are 33 stars within 12.5 lightyears (Powell, 2006), so our local stellar density can be calculated as

$\delta = \frac{N}{\pi * \frac{4}{3} * \pi * R^3} = 0.004$ stars per cubic lightyear. Using this density, we calculate that we have reached 8,531 stars in 80 years of broadcasting, and thus 3,555 earthlike planets.

Those numbers are what make the Fermi paradox so counterintuitive. We have reached *so many stars and planets,* surely we should have reached somebody by now, and in turn been reached! However, these numbers are dwarfed by the size and number of stars/planets in the Milky Way galaxy. The disk of our galaxy has a volume of $6.54 * 10^{12}$ cubic lightyears $V = \frac{4}{3} * \pi * R^3$ and contains 200 billion stars, of which roughly $8.33 * 10^{10}$ have earthlike planets (Tremaine and Binney, 1987). So humanity has reached $3.23 * 10^{-5}$ % of the volume, and $4.24 * 10^{-6}$ % of the stars that make up the Milky Way galaxy. Even our mundane, typical spiral galaxy (it is not exceptionally large compared to other galaxies) is vast beyond imagining (NASA, 2009). This is why the Fermi paradox is counterintuitive- the numbers involved are large enough that they defy intuitive human understanding. With this in mind an objective, completely non-intuitive analysis of the Fermi paradox was undertaken, to demonstrate that it is not, in fact, odd that we *appear* to be alone.

## An Explanation for the Fermi Paradox

We seem to be alone. We have not heard from any other civilizations, have not seen any or, to our knowledge, reached them. But there are *so many stars* that this seems very odd. By the mediocrity principle, we know that we are most definitely not anything special in the universe. The fact that life arose on Earth means that it should have arisen many other places, because we are in almost every way we know typical (Vilenkin, 2006). But if life is common, where is it?

The answer, it turns out, may be that we simply have not heard *yet,* and will be communicating with extraterrestrials sometime in the not-too-distant future (too distant for anybody alive today, but only thousands of years). This was determined by an analysis of what it means that we have not heard *yet*, in terms of what portion of the galaxy may have been reached and how many civilizations we predict there may be in the galaxy.

### Derivation of Inequality

To determine these predictions, an inequality was derived relating the frequency with which life arises on a suitable planet, $f_l$, and the average length of the broadcasting history of advanced civilizations (time between first broadcast and today), $L_H$. In determining this relation, the galaxy was assumed to be a flat, planar disk with radius 10kpc, containing 200 billion stars (Tremaine and Binney, 1987), $1/2.4$ of which have earthlike planets in orbit around them (Terzian and Hohlfeld, 1976). The resulting relation (found by computing the portion of the Milky Way we expect to have been reached for a given $f_l$ and $L_H$) can be used to make estimates of when we can expect to be reached, and to demonstrate that it is actually reasonable that we have not heard from anybody else yet.

The planar area in the galactic disk reached by communication from any intelligent civilization (assuming such civilizations are uniformly dispersed throughout the galaxy) can be modeled as the area of N disks with radius $L_H$ (average length of broadcasting history in years).

$$A_c = N * \pi * r_c^2, r_c = L_H$$

Frank Drake's Green Bank equation yields an estimate of the number of intelligent, communicating civilizations currently active in the Milky Way. For the purposes of this research, we derived a modified version of this equation predicting N as the number of civilizations which have ever existed in galactic history. We assumed that all life eventually evolves intelligence and the technology necessary to broadcast communication.

$$N = n_s * f_p * n_e * f_l * f_c \approx \frac{1}{2.4} * n_s * f_l \quad (f_p = f_c = 1, n_e = \frac{1}{2.4})$$

Substituting the N from the modified Drake equation into the formula for area reached:

$$A_c = \frac{1}{2.4} * n_s * f_l * \pi * L_H^2$$

Dividing this area by the total planar area of the galactic disk, we arrive at a formula for the proportion of the galaxy reached by broadcast communication. Setting this portion to less than one half (because, by the mediocrity principle, we would expect to be reached when approximately ½ of the galaxy has been reached and we have not yet been reached), we derive an inequality relating several constant and variable terms.

$$P_c = \frac{A_c}{A_G} = n_s * f_l * \pi * L_H^2 \Big/ 2.4 * A_G < \frac{1}{2}$$

Separating the variable and constant terms (variable on the left-hand-side and constant on the right), the inequality now relates the frequency of life arising on a suitable planet, the average length of broadcasting history for any capable civilization, and a constant term.

$$f_l * L_H^2 < \frac{1.2 * A_G}{\pi * n_s}$$

This inequality can be used to determine limits on the unknowns (the frequency of life arising on a suitable planet and the average length of time since the first broadcast of such a civilization), and to offer predictions determined by certain parameters.

**Calculating Inequality Constant**

Taking the galaxy as a disk with radius 32,620 light years 200 billion stars (Tremaine and Binney, 1987), the constant term in the inequality is determined to be $6.38 * 10^{-3}$ with units of $ly^2/_{star}$. We conjecture this as a constant which is time-dependent and differs between galaxies. However, the scale of time in which it changes (cosmological time) is such that for our purposes in the Milky Way it can be considered to be constant with respect to time.

**Conjecturing History Upper Bound**

An upper limit for the length of the broadcasting history of an average civilization can be arrived at through the mediocrity principle. By this principle, we know that humanity was almost certainly not the (or even *one of the*) first species in the galaxy to develop broadcasting technology, nor one of the last. Put statistically, it can be said with a high degree of confidence that humanity is somewhere in the median 90% of the population of galactic species as far as broadcasting history is concerned. That is to say, we are not among the first nor last 5% of civilizations to develop this technology.

Taking a very conservative estimate, we posit that we have been broadcasting for 5% as long as the average communicative species has been, and as such this upper limit on the average is approximately 1600 years. This can be substituted back into the inequality derived previously to give an idea of the frequency of life that our apparent loneliness suggests.

Substituting this value for $L_H$, we arrive at the bound for frequency of life $f_l \leq 2.52 * 10^{-6}$ occurrences per suitable star. Using this frequency and history length estimate in our modified Drake's equation (see derivation of inequality), we arrive at there being fewer than 210 intelligent, communicating civilizations in galactic history. By the methods typically used (primarily the unmodified Drake's equation), this is a quite reasonable number.

## Assuming Mediocrity

We now take the mediocrity principle to its extreme conclusion- that we are exactly representative of the average intelligent civilization in the galaxy, and there is nothing special about us. With this in mind, we take the average length of broadcasting history to be the same as ours, 80 years (Trawick, 2011). We use our previously-derived value for the number of civilizations (210) to determine the planar area reached so far by *any* intelligent civilization. Using the formula $A_c = \pi * R_c^2$, we estimate that $4.18 * 10^6$ square light years has been reached by some type of broadcast communication. This seems a vast area, until it is considered that this is only 0.125% of the planar area of the Milky Way galaxy. For us to have been reached so far, we would need to be in this special, proportionally tiny area of the galaxy, and we know we are not special. This calculation is thus offered as a numerical justification for the fact that we have not been reached, and as a conceptual explanation for the Fermi paradox.

## Conjecturing Frequency Upper Bound

To calculate an upper bound for the frequency of life arising on a suitable planet, we assume that we are *about to* hear a response to our first communication. In other words, the leading edge of our communication (which has been relatively constant since then) reached another civilization approximately 40 years ago, and they immediately sent a response signal to indicate having heard us. Assuming that we are the only civilization in a sphere of 40 light years, but that the closest civilization is exactly 40 light years away, we arrive at the estimate that frequency of life occurring is $9.38 * 10^{-4}$. This frequency, suggests (according to our inequality) an upper bound for the average length of broadcasting history of 3 years, and predicts 78.1 million civilizations in galactic history. This would mean that we were among the first to develop broadcasting technology, and that life is not only common, it is ubiquitous.

Both of these conceptual conclusions violate the mediocrity principle, and so we reject them. We are left with the conclusion that we are *not* about to hear back from another civilization, because if we were, we would have *reducto ad absurdum*. If we were about to hear back, we would have to be very special, and we take as an assumption that we are in no way special, so we therefore are not about to hear back.

### When Our Loneliness Becomes Unexpected

As discussed, we expect to have heard from an alien civilization when approximately half of the galaxy has been reached. Using the same values as in the "Assuming Mediocrity" section, it is calculated that $L_H$ would need to be approximately 1,580 years before half of the galaxy is reached and it becomes slightly disconcerting that we have not. Taking our broadcasting history to be the average, we thus claim that if we still have not heard anything 1,500 years from now, the Fermi paradox begins to become something other than a statistical prediction. This is not to say that we *must* be reached by then or else we are, in fact, alone. We simply claim that it is somewhat unlikely (fifty percent unlikely) that we will not hear anything until beyond that time.

### Conclusions

The calculations performed here are left primarily as evidence that the so-called Fermi paradox is not, in fact, paradoxical. It is instead a quite reasonable prediction, given the size of our galaxy. The inequality which was derived is offered as our primary evidence for this conclusion, and as a relation which we hope will be explored by others.

Any calculations which started with the assumption that we *should* have been reached already yielded utterly absurd results, and so the premise (that we should have heard by now) was also discarded as absurd. Values which assumed it to be expected that we would not have heard yielded, in turn, quite reasonable results, and so it is concluded that we should not expect to have heard from another civilization yet.

Though the Fermi paradox is undeniably counterintuitive, this paper is offered as an argument that it is not, in fact, unreasonable that we have thus far *appeared* to be alone. We may very well be reached someday. In fact, by the mediocrity principle, we should *expect* to be. But that day is not now, or any time in the immediate future.

The calculation of how long we have to wait until 50% of the galaxy has *most likely* been reached (assuming the extreme case of absolute mediocrity) suggests that we have 1,500 years until we can start to truly identify the Fermi paradox as an actual paradox. In the intervening time, we should continue looking, and continue letting the galaxy know that we are here, in the hope that someday, somebody will answer us.

## Discussion of Limits

This analysis was limited in several ways, and cannot be used to determine exact values for certain parameters or probabilities. It can, however, be used to offer estimates and a more rigorous answer to the Fermi paradox.

First, the inequality derived was just that, an inequality. With this in mind, all resulting predictions for the parameters are given as estimates and upper bounds, because the less-than relation can only offer bounds. Many of the values arrived at as initial estimates/bounds for the parameters are also offered only as bounds, for the same reason.

Second, the galaxy was assumed to be a flat disk, and communication spheres were assumed to have no overlap. If there were overlap, then it is possible (even likely) that other civilizations are already in contact with each other, and by the mediocrity principle we would expect to have been contacted as well. The Milky Way was taken to be flat primarily to simplify the inequality, and justified because our understanding of the exact shape and size of the galaxy, and the distribution of stars and planets resulting from this shape, are unknown. We could not have reached a much higher level of accuracy by using a more complicated geometric model, and the inequality is offered only as a way to produce estimates, so exact values were not necessary.

These limits render invalid any exact predictions which could be made using the inequality and other information presented here (for instance, "we should hear in the year 2075" would be a completely unreasonable claim even if it were supported by the inequality). Such conjectures are offered only as approximations and bounds, and meant to provide a fairly wide range of values we would consider reasonable.

Despite these limits, we believe the inequality derived in this paper has merit as a tool to produce such estimated ranges, and the conclusions reached (almost strictly conceptual ones) are valid. Many approximations are used, but they were used in pursuit of an approximate answer to a previously-unanswered question, and so we are satisfied with the level of precision our calculations are capable of providing.

## Limits of Technology

The analysis of our broadcasting history and the probabilistic account of how justified we are in *expecting* to be reached both assume that any civilization (including our own) which is reached by the radio signals of another civilization would be able to receive them. As stated in the derivation of the inequality, it is assumed that any planet which develops life would eventually develop intelligence, and in turn broadcasting/receiving technology. Even with these optimistic assumptions, there is another obstacle for any dialog between civilizations to overcome: The intensity of a radio broadcast decreases as the inverse square of distance traveled, and for interstellar distances this has the potential to diminish all signals enough that they become unnoticeable. How powerful would a signal need to be in order to be realistically receivable?

Because signal strength decreases as the inverse square of distance, the power received by an observer (who must be observing the correct wavelength band) can be easily calculated.

$$P_o = P_B * \frac{R^2}{4 * d^2}$$

Where $P_o$ is power observed, $P_B$ is power broadcast, R is the radius of the receiving radio telescope, and d is the distance separating the broadcaster and the receiver. Each of these parameters can change from scenario to scenario, but $P_o$ must be at a minimum level in order to be heard. This level varies depending on many factors, especially level of technological sophistication. To come up with a lower limit on what signals can be heard we use one of the weakest signals that we receive here on Earth: The data transmissions from Voyager I. This probe, launched in 1977, is now 134 AU from Earth, and still broadcasting data back to us with its 22 Watt antennae. These signals are received by dishes like the JPL's 70-meter Deep Space Station at Goldstone, CA. The signals arrive and register a power on the dish of $1.5*10^{-23}$ Watts (Poon, 2006). We use this total received power as our estimate of the weakest signals that a civilization could be reasonably expected to receive.

**How Far Can We Communicate?**

Cornell University's Arecibo telescope is the most sensitive astronomical instrument ever developed on Earth (Reynolds, 2010). It is capable of observing signals that register that minimum power across its 152.5 m radius dish from sources significantly weaker/farther than Voyager I. If the average communicative civilization can observe signals which arrive at their planets with this strength, and they have Arecibo-sized radio telescopes, then any such civilizations within a given distance d should be able to hear a broadcast with initial power $P_B$.

$$d = \sqrt{3.87 * 10^{26} * P_B}$$, with d in units of meters

A signal strong enough to be heard by any civilizations within of our local stellar neighborhood (60 lightyears) would need to be broadcast with $8.33 * 10^8$ Watts of power.

Fortunately, our species has sent out a signal far more powerful than this, and is listening hard enough to hear a similar signal from another planet. Cornell's Arecibo telescope is not only our most powerful observational tool, it is also our species' most powerful transmitter. In 1974, Arecibo made the most powerful broadcast to ever leave our planet. The message was binary code displaying a hieroglyphics-like message graphically showing some identifying characteristics of our planet and species. Because no antennae has ever been made powerful enough to send this signal out in a sphere with sufficient intensity, Arecibo made many individual, focused broadcasts that, when considered together, are equivalent to one 20-terawatt, omnidirectional broadcast. This is powerful enough to eventually reach every planet in the galaxy with at least the minimum power described above. With a broadcast that powerful, we predict Arecibo's signals will be receivable 9,232 lightyears away. For reference, the Milky Way's disk has a radius of approximately 32,600 lightyears. However, by the time the Arecibo message has reached that distance, the sophistication of the technology of ETI will have advanced considerably, making their minimum received power much lower and the effective radius of communication much higher.

However, this signal has only reached approximately 1,263 stars, and thus 526 earthlike planets. These numbers are insignificant portions of the Milky Way ($6.32*10^{-7}$ %), and so we have quite some time to wait until we can expect to have reached another civilization with our broadcasts. With this being our *effective* broadcasting range so far, we would not expect to have reached another civilization yet (or been reached, again assuming extreme mediocrity)unless there were 158 million civilizations evenly dispersed throughout the galaxy. This is a completely absurd consequence, and goes to further demonstrate the conclusions of the probabilistic model.

## Acknowledgments

The authors thank Cornell University for providing access to its resources and faculty, and for its continued support of astronomy research. The NASA Space Grant program in the state of New York, with director Mason Peck and associate director Erica Miles, provided financial support for the project, and Erica Miles in particular provided logistical support and advice. All resources cited here are thanked for doing significant work either researching or publicizing space science, without which further research would be impossible.

Researcher Evan Solomonides thanks his parents for their continued support of his education and ambitions, Maria Yampolskaya for her enthusiasm about this project, and Lisa Kaltenegger and all of the guest lecturers in ASTRO 2299 in the fall 2015 semester for encouraging inquisitive thinking about the search for life in the universe. He also would like to thank Patricia Fernandez de Castro for helping him begin this project in such a significant capacity and in helping see it through to its completion, and his coauthor for providing access to his wealth of astronomical information, his willingness to undertake such a project with an undergraduate, and the access he provided to the astronomy department and its resources.

Researcher Yervant Terzian would like to also thank Shami Chatterjee, Jim Cordes, and Frank Drake for valuable contributions to the field and this paper.